\documentclass[pra,twocolumn,showpacs,preprintnumbers,amsmath,amssymb,notitlepage]{revtex4-1}

\usepackage{graphicx}
\usepackage{dcolumn}
\usepackage{bm}
\usepackage{comment}
\usepackage{bbold}
\usepackage{amsmath}

\begin{document}

\title{Floquet engineering from long-range to short-range interactions}
\date{\today}

\author{Tony E. Lee}
\affiliation{Department of Physics, Indiana University-Purdue University Indianapolis (IUPUI), Indianapolis, Indiana 46202, USA}

\begin{abstract}
Quantum simulators based on atoms or molecules often have long-range interactions due to dipolar or Coulomb interactions. We present a method based on Floquet engineering to turn a long-range interaction into a short-range one. By modulating a magnetic-field gradient with one or a few frequencies, one reshapes the interaction profile, such that the system behaves as if it only had nearest-neighbor interactions. Our approach works in both one and two dimensions and for both spin-1/2 and spin-1 systems. It does not require individual addressing, and is applicable to all experimental systems with long-range interactions: trapped ions, polar molecules, Rydberg atoms, nitrogen-vacancy centers, and cavity QED. Our approach allows one achieve a short-range interaction without relying on Hubbard superexchange.
\end{abstract}

\maketitle

\section{Introduction}

A quantum simulator is a quantum system that is engineered to implement a particular quantum model \cite{hauke12,georgescu14}. A quantum simulator with a large number of particles would be able to simulate quantum many-body systems beyond what a classical computer could handle \cite{feynman82}. One goal of quantum simulation is to implement models that describe solid-state systems, and thereby gain direct insight into phenomena like high-$T_c$ superconductivity. 

There has been a lot of progress on quantum simulation using cold atoms \cite{hauke12,georgescu14}. A common feature of such systems is the presence of long-range interactions that decay with a power law in distance due to dipolar or Coulomb interactions \cite{porras04,gorshkov11,bhongale13,saffman10,cai13,morrison08}. On the one hand, long-range interactions can lead to qualitatively new physics \cite{maghrebi15}. On the other hand, solid-state systems usually have short-range interactions because Wannier functions are exponentially localized \cite{kohn59,brouder07,ashcroft78}. Thus, for the sake of directly simulating solid-state models, it can be preferable for quantum simulators to have short-range interactions.

For ultracold atoms in an optical lattice, the on-site interaction arising from $s$-wave scattering allows one, in principle, to achieve a nearest-neighbor spin model via superexchange \cite{duan03,bloch08}. However, the nearest-neighbor interaction is small, and it is hard to cool the atoms to sufficiently low temperatures. This has motivated many experimental groups to create quantum simulators based on dipolar or Coulomb interactions \cite{porras04,gorshkov11,bhongale13,saffman10,cai13,morrison08}. The advantages of these setups are that the interaction strength is large and that the atoms do not have to be very cold. However, these setups have long-range interactions, so it would be beneficial to somehow remove the long-range tail while otherwise preserving the magnitude of the interactions.

\begin{figure}[b]
\centering
\includegraphics[width=3.4 in,trim=1.1in 4.3in 1.2in 4.2in,clip]{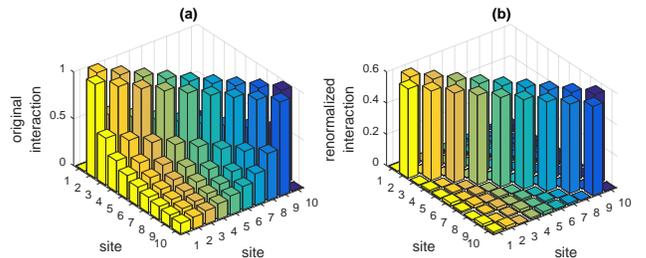}
\caption{\label{fig:Jmatrix_alpha1}Interaction profile, showing interaction strength between spins of a one-dimensional chain with 10 spins. (a) Original interactions that decay in distance as $1/r$. (b) Renormalized interactions $|\beta_r|/r$ due to modulation of the field gradient with $N=3$ frequencies with $\vec{g}=(0.640,-0.377,-1.226)$. The interaction is in units of $J$.}
\end{figure}

In this Rapid Communication, we show how to use Floquet engineering \cite{bukov15,eckardt16} to reshape a long-range interaction into a short-range one. Although we focus on making the interaction as short range as possible, our approach can be used to engineer other interaction profiles. Starting from a spin model with long-range $XX$ interactions, we modulate a magnetic-field gradient periodically in time, so that in a rotating frame, the interaction profile is effectively short range. Our approach works in both one and two dimensions and for both spin-1/2 and spin-1 systems. An example result in one dimension is shown in Fig.~\ref{fig:Jmatrix_alpha1}. 

Our approach is related to the phenomenon of ``dynamical localization,'' where periodically modulating a system effectively suppresses the nearest-neighbor interaction \cite{dunlap86,lignier07}. Here, we modulate the system to suppress all interactions \emph{except} the nearest-neighbor interaction. 

Previous works have proposed schemes to suppress long-range interactions by individually addressing each spin \cite{korenbilt12,yao12}. The advantage of our approach is that it does not require individual addressing, since the magnetic-field gradient acts on all spins at the same time. Also, our approach is universal and can be applied to all experimental systems with long-range interactions: trapped ions \cite{porras04}, polar molecules \cite{gorshkov11,bhongale13}, Rydberg atoms \cite{saffman10}, nitrogen-vacancy (NV) centers \cite{cai13}, and cavity QED \cite{morrison08}.

We first discuss the one-dimensional (1D) case and then the two-dimensional (2D) case. For one dimension, we present three different schemes in order of increasing complexity but increasing potency.

\section{1D model}

Consider a one-dimensional spin chain with long-range $XX$ interactions. We modulate the system with a time-dependent gradient in the transverse field,
\begin{eqnarray}
H(t)&=&\sum_n \sum_{r\geq1} \frac{J}{r^\alpha} \left(\sigma^x_n \sigma^x_{n+r} + \sigma^y_n \sigma^y_{n+r}\right) + \frac{\Omega}{2}\sum_n f_n(t) \sigma^z_n, \nonumber\\ \label{eq:H_1d}
\end{eqnarray}
where $f_n(t)$ is a periodic function with frequency $\Omega$ and period $T$. We let the gradient strength scale with $\Omega$ in order to get a nontrivial Floquet Hamiltonian in the limit of large $\Omega$ \cite{bukov15,eckardt16}. The gradient can be generated experimentally by a magnetic field \cite{jotzu15,luo16} or ac Stark shift \cite{aidelsburger13,miyake13}. We assume power-law interactions, where $\alpha$ is the exponent. In this Rapid Communication, we focus on $\alpha=3$, which is relevant to the common situation of dipolar interactions. 

We go into the interaction picture, rotating with the last term in Eq.~\eqref{eq:H_1d}. The wave function in the rotating frame $|\psi'\rangle$ is related to that in the laboratory frame $|\psi\rangle$ via
\begin{eqnarray}
|\psi'(t)\rangle&=&U^\dagger (t)|\psi(t)\rangle,\\
U(t)&=&\exp\left[-i\int_0^t dt' \frac{\Omega}{2}\sum_n f_n(t') \sigma^z_n \right]. \label{eq:U}
\end{eqnarray}
In the rotating frame, the Hamiltonian is
\begin{eqnarray}
H'(t)&=&\sum_n \sum_{r\geq1} \frac{J}{r^\alpha} \cos\left[\int_0^t dt' \Omega [f_{n+r}(t')-f_n(t')]\right] \nonumber\\
&& \quad\quad\quad\quad\times\left(\sigma^x_n \sigma^x_{n+r} + \sigma^y_n \sigma^y_{n+r}\right). \label{eq:Hr_1d}
\end{eqnarray}
Note that Eq.~\eqref{eq:Hr_1d} is still exact --- we have only applied a unitary transformation. Also, $f_n(t)$ should be of a form, such that $H'(t)$ is periodic and we can then apply Floquet theory.

Now we assume that the modulation frequency $\Omega$ is large: $\Omega\gg J$. In this limit, we can make a rotating-wave approximation: according to Floquet theory \cite{bukov15,eckardt16}, the lowest-order Hamiltonian is obtained by averaging Eq.~\eqref{eq:Hr_1d} over one period of modulation, leading to the time-independent Floquet Hamiltonian $H_F$:
\begin{eqnarray}
H_F&=&\sum_n \sum_{r\geq1}  \frac{J\beta_r}{r^\alpha}\left(\sigma^x_n \sigma^x_{n+r} + \sigma^y_n \sigma^y_{n+r}\right), \label{eq:HF_1d} \\
\beta_r &=& \frac{1}{T} \int_0^T dt \cos\left[\int_0^t dt' \Omega [f_{n+r}(t')-f_n(t')]\right].
\end{eqnarray}
So we have an $XX$ chain where the interaction strength between two spins is renormalized by a distance-dependent factor $\beta_r$. Thus, the interaction profile can be shaped via $f_n(t)$. To get only nearest-neighbor interactions, we choose $f_n(t)$ so as to suppress all $\beta_r$ except for $\beta_1$. 

Note that $|\beta_r|\leq 1$, so the renormalized interaction is always smaller than the original interaction. Also, in this Rapid Communication, $\beta_r$ is always real, independent of $n$, and even in $r$.

The above derivation applies to all spin magnitudes, not just spin-1/2. This means we can similarly shape the interaction profile of a long-range spin-1 chain. This is relevant because there are several proposals for implementing spin-1 models in atomic or molecular systems \cite{brennen07,gorshkov13,vanbijnen15,gong16a}.

The wave functions in the rotating and laboratory frames are related by the unitary transformation $U(t)$ in Eq.~\eqref{eq:U}, so at the end of the experiment (after evolving $|\psi\rangle$ with $H$ for time $t$), one has to apply $U^\dagger(t)$ to convert from the laboratory frame to the rotating frame. However, $U(t)$ is usually very simple when $t=mT$, where $m$ is an integer. For example, for $f_n(t)$ defined in Eqs.~\eqref{eq:f_single} and \eqref{eq:f_multi}, $U(mT)=1$ and $|\psi'(mT)\rangle=|\psi(mT)\rangle$, so no transformation is needed at these stroboscopic times.

%We try to find $f(t)$ such that the interaction $J\beta_r/r^\alpha$ is as close to nearest-neighbor as possible.

In general, $\beta_r$ does not have a simple form, so we need a way to quantify how short range the interaction is. We use the quantity
\begin{eqnarray}
\delta^2 &=& 2\sum_{r=2}^{M-1} \left(\frac{\beta_r}{\beta_1 r^\alpha}\right)^2 \label{eq:delta}
\end{eqnarray}
to estimate how close the Floquet Hamiltonian [Eq.~\eqref{eq:HF_1d}] is to a perfect nearest-neighbor model. $\delta$ is a rough estimate of the rate at which the evolution of $H_F$ deviates from a nearest-neighbor model for a chain of $M$ spins. The reason is that, in the time scale set by the nearest-neighbor interaction $J\beta_1$, the longer-range interactions are $\beta_r/(\beta_1 r^\alpha)$. Assuming that any population that evolves via non-nearest-neighbor interactions is lost forever, $\delta$ is roughly the rate at which population leaks out.

A perfect nearest-neighbor interaction would have \mbox{$\delta=0$}. For a long chain with $\alpha=3$ and without renormalization ($\beta_r=1$), $\delta=0.186$. One could define $\delta$ differently to identify other interaction profiles, e.g., allowing next-nearest-neighbor interactions.

\subsection{Linear gradient: One frequency}

First, we consider a linear gradient that includes a static component and a single frequency, 
\begin{eqnarray}
f_n(t)&=& n[-g_0 + g_1 \cos(\Omega t)], \label{eq:f_single}
\end{eqnarray}
where $g_0$ is assumed to be an integer so that the static gradient is resonant with $\Omega$. In the rotating frame,
\begin{eqnarray}
H'(t)&=&\sum_n \sum_{r\geq1} \frac{J}{r^\alpha} \cos\left[-\Omega r g_0 t + r g_1 \sin(\Omega t)\right] \nonumber\\
&& \quad\quad\quad\quad\times\left(\sigma^x_n \sigma^x_{n+r} + \sigma^y_n \sigma^y_{n+r}\right). \label{eq:Hr_1d_one}
\end{eqnarray}
Note that the unitary transformation $U(t)$ includes both the static and dynamical components of the gradient. Since $g_0$ is an integer, $H'(t)$ is periodic in time, so Floquet theory can be applied.
Then after taking the rotating-wave approximation, we obtain the Floquet Hamiltonian $H_F$ in Eq.~\eqref{eq:HF_1d} with renormalized interactions,
\begin{eqnarray}
\beta_r &=& \frac{1}{T} \int_0^T dt \cos\left[-\Omega r g_0 t + r g_1 \sin(\Omega t)\right] \\
&=& \mathcal{J}_{rg_0}(rg_1), \label{eq:beta_field_grad_one}
\end{eqnarray}
where $\mathcal{J}_n(z)$ is the Bessel function of the first kind.

We mention a few special cases with $g_0=0$ to illustrate the shaping of the interaction. Using the asymptotic form of the Bessel function \cite{abramowitz64},
\begin{eqnarray}
g_1=2\pi \quad &\rightarrow& \quad \beta_r\approx \frac{1}{\pi\sqrt{2r}}, \label{eq:g1_2pi}\\
g_1=\pi \quad &\rightarrow& \quad \beta_r\approx \frac{(-1)^r}{\pi\sqrt{r}}, \label{eq:g1_pi}\\
g_1=\frac{2\pi}{\ell} \quad &\rightarrow& \quad \beta_r\approx \frac{1}{\pi}\sqrt{\frac{\ell}{r}}\cos\left(\frac{2\pi r}{\ell}-\frac{\pi}{4}\right). \label{eq:g1_2pi_ell}
\end{eqnarray}
Equation \eqref{eq:g1_2pi} means that the power-law exponent is increased by 1/2. Equation \eqref{eq:g1_pi} means that the interaction alternates sign; this can be used to stabilize an antiferromagnetic phase \cite{gong16a}. Equation \eqref{eq:g1_2pi_ell} means that the interaction is modulated in distance with wavelength $\ell$; this can be used to stabilize a spin-density-wave phase.

\begin{figure}[t]
\centering
\includegraphics[width=3.6 in,trim=1.1in 4.1in 1in 4.1in,clip]{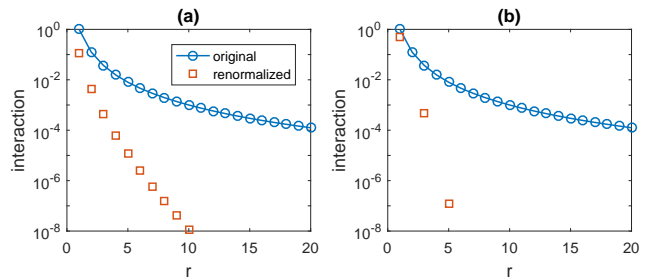}
\caption{\label{fig:1d_single}Original ($1/r^3$) and renormalized ($|\beta_r|/r^3$) interactions for a 1D chain, modulated in two different ways. (a) Modulated field gradient with $g_0=2,g_1=1$. (b) Running lattice with $g_0=2,g_1=3.05,\phi=\pi$. The interaction is in units of $J$.}
\end{figure}

It turns out that when $g_0>g_1$, $\beta_r$ decays exponentially with $r$. This is seen from the asymptotic form of Eq.~\eqref{eq:beta_field_grad_one} for large $r$ \cite{abramowitz64}:
\begin{eqnarray}
\beta_r&\approx& C e^{-r/\ell},\\
1/\ell &=& g_0\,\text{sech}^{-1}\left(\frac{g_1}{g_0}\right) - \sqrt{g_0^2-g_1^2},\\
1/C &=& \sqrt{2\pi r\sqrt{g_0^2-g_1^2}}.
\end{eqnarray}
Thus, we have already achieved a short-range interaction. As $g_0$ increases, $\ell$ decreases (shorter range). Figure \ref{fig:1d_single}(a) shows that the renormalized interaction $\beta_r/r^\alpha$ decays exponentially with $r$; for the example shown with $\alpha=3$, $\delta=0.053$ and $\beta_1=0.11$.

\begin{figure}[b]
\centering
\includegraphics[width=3.4 in,trim=0.2in 2.2in 0.2in 1.5in,clip]{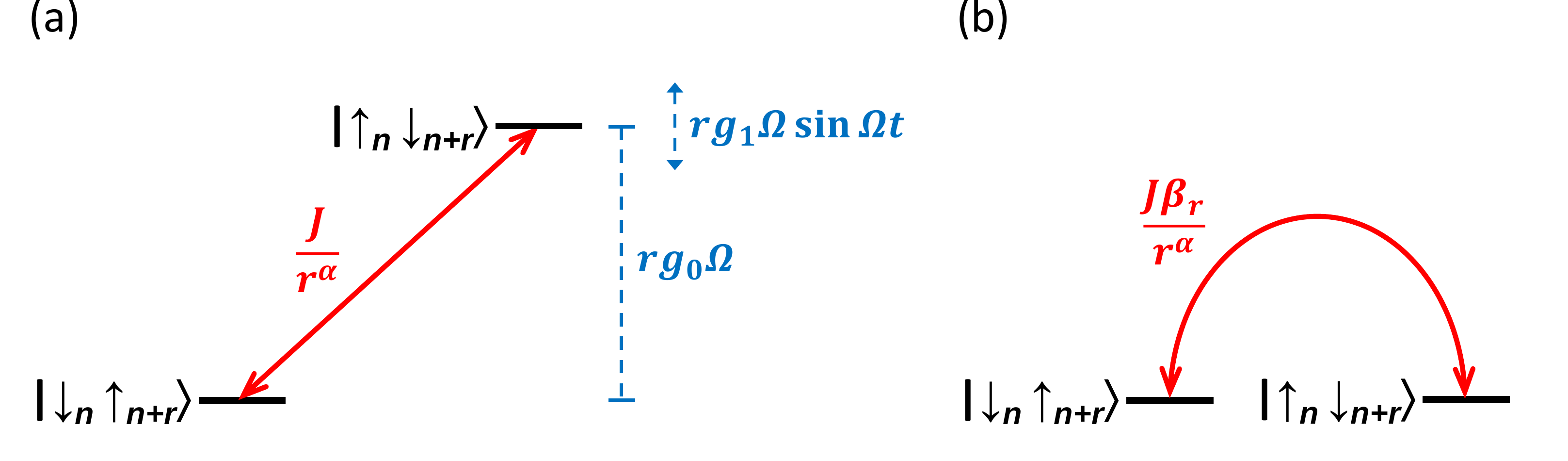}
\caption{\label{fig:1d_diagram}Illustration of energy levels with modulated field gradient in (a) laboratory frame and (b) rotating frame.}
\end{figure}

The exponential decay can be intuitively understood as follows (see Fig.~\ref{fig:1d_diagram}). The term $\sigma^x_n \sigma^x_{n+r} + \sigma^y_n \sigma^y_{n+r}$ in $H$ causes a transition between $\left|\downarrow_{n}\uparrow_{n+r}\right\rangle$ and $\left|\uparrow_{n}\downarrow_{n+r}\right\rangle$. Due to the gradient, the frequency detuning of these two states includes a static component $rg_0\Omega$ and a time-dependent component $rg_1\Omega\sin\Omega t$. (This is essentially a many-body generalization of the Rabi model, i.e., a two-level system with a periodic drive \cite{ashhab07,zhou09,lee15}.) Thus, the transition is an $rg_0$-photon transition; i.e., to undergo the transition, the system has to absorb $rg_0$ ``photons'' from the periodic drive. When $g_0>g_1$, the driving strength is weak relative to the static detuning. Then as $r$ increases, the transition probability decreases exponentially because it is a higher-order transition. In the rotating frame [Fig.~\ref{fig:1d_diagram}(b)], this means that $\beta_r$ decreases exponentially with $r$.

Although this $f_n(t)$ leads to exponentially decaying interactions, it is not suitable for generating only nearest-neighbor interactions. The reason is that as $\ell$ decreases, $\beta_1$ also decreases. So if we set $g_0$ large to have only nearest-neighbor interactions, the strength of that interaction would be small. This would be problematic in practice, because one would have to run the experiment for a long time, and the evolution would be dominated by decoherence. Ideally, we would have only nearest-neighbor interactions with $\beta_1$ on the order of unity. To get around this issue, we next discuss multiple frequencies.

\subsection{Linear gradient: Multiple frequencies}

Now we modulate the linear gradient with $N$ different frequencies (harmonics of $\Omega$),
\begin{eqnarray}
f_n(t)&=& n \sum_{k=1}^N k g_k \cos(k\Omega t), \label{eq:f_multi}
\end{eqnarray}
where $kg_k$ is the amplitude of the $k$th harmonic. For simplicity, we omit the constant term $(g_0=0)$, so there are $N$ different parameters to tune. Then $\beta_r$ is
\begin{eqnarray}
\beta_r &=& \frac{1}{2\pi} \int_0^{2\pi} dt \cos\left[r\sum_{k=1}^N g_k \sin(kt)\right], \label{eq:beta_field_grad_multi} 
\end{eqnarray}
which is a multidimensional Bessel function that can be more efficiently calculated as a discrete sum \cite{korsch06,verdeny15}.

%\beta_r &=& \mathcal{J}_0^{\vec{k}} (r\vec{g}). \\

\begin{figure}[t]
\centering
\includegraphics[width=3.6 in,trim=1.1in 4.1in 1in 4.1in,clip]{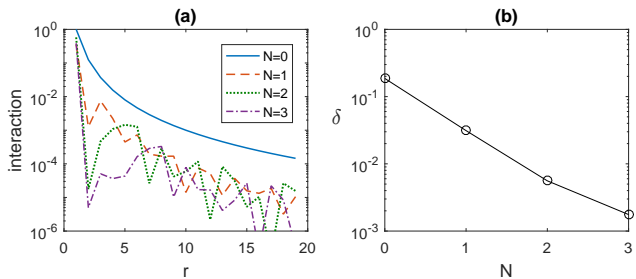}
\caption{\label{fig:1d_multi}One-dimensional chain with field gradient modulated by $N$ frequencies. (a) Renormalized interaction $|\beta_r|/r^3$ for different $N$. (b) Deviation from nearest-neighbor model. For $N=1,2,3$, we use $\vec{g}=(4.308), (0.721,1.218), (3.481,-3.762,2.815)$. The interaction is in units of $J$.}
\end{figure}

We want to set $g_k$ so as to make the renormalized interaction, $J\beta_r/r^\alpha$, as short range as possible. In other words, the goal is to minimize $\delta$ in Eq.~\eqref{eq:delta} with respect to $g_k$. Since $\delta$ is a complicated function of $g_k$, we use a quasi-Newton algorithm to search for the optimal values of $g_k$. An example for $\alpha=1$ is shown in Fig.~\ref{fig:Jmatrix_alpha1}. Examples for $\alpha=3$ are shown in Fig.~\ref{fig:1d_multi}. As $N$ increases, the long-range interactions are more suppressed and $\delta$ decreases. 

The suppression of long-range interactions comes at the cost of a reduced nearest-neighbor interaction \mbox{($\beta_1<1$)}. For the examples shown in Fig.~\ref{fig:1d_multi}(a), $\beta_1\sim 0.5$. Thus, we have succeeded in generating a nearest-neighbor model with a relatively large interaction strength. If desired, one could use values of $g_k$ that are slightly less optimal in terms of $\delta$ but have larger $\beta_1$.

\subsection{Running lattice}

We now discuss another scheme that is more powerful than the ones discussed above. We modulate the chain with a running lattice but also include a static gradient,
\begin{eqnarray}
f_n(t)&=&  n g_0 + \frac{g_1}{2} \sin(\Omega t - \phi n),
\end{eqnarray}
where $g_0$ (an integer) is the strength of the static gradient, while $g_1$ is the amplitude of the running lattice. The running lattice can be generated experimentally by interfering two laser beams at slightly different frequencies \cite{aidelsburger13,miyake13}. The phase $\phi$ is determined by the wavelength of the running lattice and the spin separation. 

%we use only one modulation frequency, although multiple frequencies would lead to even better results.

%There are three parameters ($g_0,g_1,\phi$), which we use to minimize $\delta$.

%First consider the simplest case of no static gradient ($g_0=0$). In the interaction picture and in the limit of large $\Omega$, the Floquet Hamiltonian is Eq.~\eqref{eq:HF_1d_field_grad} with
%\begin{eqnarray}
%\beta_r &=& \mathcal{J}_{0}\left[g_1\sin\left(\frac{r\phi}{2}\right)\right]. \label{eq:beta_running_nograd}
%\end{eqnarray}
%By suitably choosing $g_1,\phi$, one can achieve a fairly low $\delta\approx 0.013$ for $\alpha=3$.

%Next, we include the static gradient ($g_0\neq 0$). 
%To simplify the discussion,  Then

For simplicity, we assume that $g_0$ is an even integer and $\phi=\pi$. Then the Floquet Hamiltonian [Eq.~\eqref{eq:HF_1d}] has renormalized interactions with
\begin{eqnarray}
\beta_r &=& (-1)^\frac{g_0}{2} \mathcal{J}_{rg_0}\left[g_1\sin\left(\frac{r\pi}{2}\right)\right] \quad\text{for odd $r$}, \label{eq:beta_running_grad}
\end{eqnarray}
and $\beta_r=0$ for even $r$. An example is shown in Fig.~\ref{fig:1d_single}(b). We find that $\beta_r$ decreases exponentially with $r$: \mbox{$\beta_r\sim (-1)^{g_0/2} (eg_1/2g_0r)^{g_0r}$}. In fact, all $\beta_r$ except $\beta_1$ are suppressed so much that there are essentially only nearest-neighbor interactions. But the advantage of Eq.~\eqref{eq:beta_running_grad} over Eq.~\eqref{eq:beta_field_grad_one} is that $\beta_1$ can be kept on the order of unity by suitably choosing $g_1$ \footnote{The reason is that the argument of the Bessel function increases monotonically with $r$ in Eq.~\eqref{eq:beta_field_grad_one} but is bounded in Eq.~\eqref{eq:beta_running_grad}}. In this way, one obtains a nearest-neighbor model with a relatively large interaction strength. For the example shown in Fig.~\ref{fig:1d_single}(b), $\delta=0.0013$ and $\beta_1=0.49$, which is much better than Fig.~\ref{fig:1d_single}(a). As $g_0$ increases, $\delta$ decreases exponentially.

This is a very strong scheme, because one can get arbitrarily close to a nearest-neighbor model using only a single frequency. This would be particularly useful when the original interactions are very long range ($\alpha\approx0$, which is the case for atoms coupled via a cavity \cite{morrison08}), when the other discussed schemes might not be as effective.

We note without proof that this running-lattice scheme can also suppress long-range Ising interactions (not just $XX$ interactions).

\section{2D model}

Our Floquet approach also works in higher dimensions. Here, we discuss a scheme for a 2D square lattice, although it can be extended to other lattice topologies. In general, two dimensions are more difficult than one dimension in terms of suppressing long-range interactions (more frequencies are needed to achieve the same $\delta$), so the gradient configuration needs to be chosen judiciously. 

\begin{figure}[t]
\centering
\includegraphics[width=3.3 in,trim=1.9in 3.4in 2in 3.5in,clip]{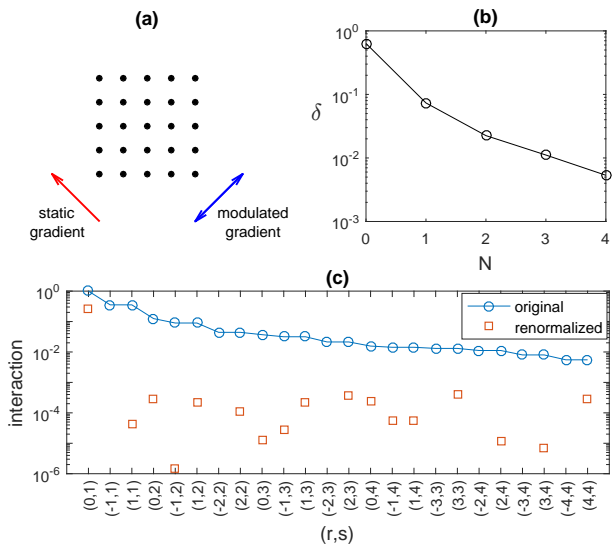}
\caption{\label{fig:2d_multi}Two-dimensional lattice of $5\times5$ spins with field gradient modulated by $N$ frequencies. (a) Diagram showing directions of the static and modulated gradients. (b) Deviation from nearest-neighbor model. (c) Original [$1/(r^2+s^2)^{3/2}$] and renormalized interactions [$|\beta_{r,s}|/(r^2+s^2)^{3/2}$] for $N=4$ with $g_0=20$ and $\vec{g}=(5.464,-3.136,5.323,1.045)$. Only some $(r,s)$ are shown. The interaction is in units of $J$.}
\end{figure}

We let the lattice have a static gradient as well as a modulated gradient, but in different directions:
\begin{eqnarray}
H(t) &=& \sum_{\substack{m,n\\r,s}} \frac{J}{(r^2+s^2)^{\alpha/2}} \left(\sigma^x_{m,n} \sigma^x_{m+r,n+s} + \sigma^y_{m,n} \sigma^y_{m+r,n+s} \right) \nonumber\\ && + \frac{\Omega}{2}\sum_{m,n}f_{m,n}(t) \sigma^z_{m,n}, \label{eq:H_2d}
\end{eqnarray}
\begin{eqnarray}
f_{m,n}(t) &=& -(m-n)g_0 + (m+n) \sum_{\substack{k=1\\ \text{odd }k}}^{2N-1} k g_k \cos(k\Omega t).\quad\quad
\end{eqnarray}
The static and modulated gradients are along diagonals of the lattice, but perpendicular to each other [Fig.~\ref{fig:2d_multi}(a)]. We assume that $g_0$ is an even integer and that the modulation includes $N$ \textit{odd} harmonics of $\Omega$. (These assumptions lead to special properties discussed below.) As before, $\alpha$ is the exponent of the power-law interaction.

%The absence of a data point means the interaction is 0.

After going into the interaction picture, rotating with the second line of Eq.~\eqref{eq:H_2d}, and taking the rotating-wave approximation, the Floquet Hamiltonian is
\begin{eqnarray}
H_F &=& \sum_{\substack{m,n\\r,s}} \frac{J\beta_{r,s}}{(r^2+s^2)^{\alpha/2}} \left(\sigma^x_{m,n} \sigma^x_{m+r,n+s} + \sigma^y_{m,n} \sigma^y_{m+r,n+s} \right), \nonumber\\
\end{eqnarray}
where the interactions are renormalized by a displacement-dependent factor,
\begin{eqnarray}
\beta_{r,s} &=& \frac{1}{2\pi} \int_0^{2\pi} dt \cos\bigg[-(r-s)g_0 t \nonumber\\ && \quad\quad\quad\quad\quad\quad\quad + (r+s)\sum_{\substack{k=1\\ \text{odd }k}}^{2N-1} g_k \sin(kt)\bigg],\quad\quad
\end{eqnarray}
which is a multidimensional Bessel function \cite{korsch06,verdeny15}. The goal now is to choose $g_k$ so that the interaction is as short range as possible, i.e., suppressing all but $\beta_{\pm1,0}$ and $\beta_{0,\pm1}$.

% \beta_{r,s} &=& \mathcal{J}^{\vec{k}}_{(r-s)g_0} \left[(r+s)\vec{g}\right]. \\

Due to the above assumptions, $\beta_{r,s}$ satisfies
\begin{eqnarray}
\beta_{r,s}=\beta_{s,r}=\beta_{-r,-s}=\beta_{-s,-r},\quad\quad \beta_{r,-r}=0. \label{eq:beta_properties}
\end{eqnarray}
These properties significantly reduce the number of independent $\beta_{r,s}$ that need to be suppressed, which reduces the number of frequencies that need to be used. For an $M\times M$ lattice, there are $M^2-M-1$ independent $\beta_{r,s}$ that need to be suppressed. Note that $\beta_{\pm1,0}$ and $\beta_{0,\pm1}$ are all equal.

To estimate how close the renormalized interactions are to a nearest-neighbor model, we use
\begin{eqnarray}
\delta^2 &=&  \sideset{}{'}\sum_{r,s=-M+1}^{M-1} \left[\frac{\beta_{r,s}}{\beta_{1,0} (r^2+s^2)^{\alpha/2}}\right]^2, \label{eq:delta_2d}
\end{eqnarray}
where $\sum'$ means to omit $(r,s)=(\pm1,0),(0,\pm1)$ from the sum. $\delta$ is again a rough estimate of the rate at which the evolution of $H_F$ deviates from a nearest-neighbor model on an $M\times M$ lattice. 
%The goal is to minimize $\delta$ with respect to $g_k$.

We again use a quasi-Newton algorithm to find the values of $g_k$ that minimize $\delta$. An example of renormalized interactions for $N=4$ and $M=5$ is shown in Fig.~\ref{fig:2d_multi}(c), where the suppression of longer-ranged interactions is evident. As the number of frequencies increases, $\delta$ decreases [Fig.~\ref{fig:2d_multi}(b)].

%We note that $\delta$ can be further decreased by using larger values for $g_0$ and $g_k$, but we have used moderate values in order to be conservative from an experimental point of view.

\section{Experimental considerations}

Our Floquet approach can be applied to all the quantum simulators with long-range interactions: trapped ions $(\alpha=\text{0--3})$ \cite{porras04}, polar molecules $(\alpha=3,5)$ \cite{gorshkov11,bhongale13}, Rydberg atoms $(\alpha=3,6)$ \cite{saffman10}, NV centers $(\alpha=3)$ \cite{cai13}, and cavity QED $(\alpha=0)$ \cite{morrison08}. The gradients can be generated using magnetic fields \cite{jotzu15,luo16} or ac Stark shifts \cite{aidelsburger13,miyake13}. We note that our approach is not limited to spin-1/2, but works equally well for spin-1 or higher \cite{brennen07,gorshkov13,vanbijnen15,gong16a}.

A potential issue with all Floquet approaches is whether the Floquet Hamiltonian is a valid description of the dynamics \cite{bukov15,eckardt16}. In deriving Eq.~\eqref{eq:HF_1d}, we assumed that the modulation frequency is large ($\Omega\gg J$), and we retained only the lowest-order term of the Magnus expansion of Eq.~\eqref{eq:Hr_1d}. If $\Omega$ is not sufficiently large, higher-order terms become relevant, causing the system to heat up from the ground state. Fortunately, spin systems are less susceptible to heating than Bose-Hubbard systems. It was shown that for spin models with two-body interactions, the heating rate decreases exponentially with the modulation frequency \cite{abanin15}; this is true even for long-range interactions \cite{mori16}. In fact, for spin models, the finite truncation of the Magnus expansion is a good approximation for times exponential in the modulation frequency \cite{kuwahara16}. In the Supplemental Material, we numerically check the accuracy of the zeroth-order Floquet Hamiltonian.

In addition, $\Omega$ should be far off-resonant from all other frequencies in the system (such as trap frequency or band gap of an optical lattice), or else the dynamics will not be limited to the Floquet Hamiltonian \cite{eckardt16}. Thus, it is better if the modulation has fewer frequencies and smaller amplitudes. But again, spin systems are less susceptible to this type of heating compared to Bose-Hubbard systems, because this heating affects motional temperature instead of spin temperature. For example, a quantum simulator based on polar molecules uses the rotational degree of freedom of the molecules to encode the spin \cite{gorshkov11}; motional heating does not significantly affect the spin dynamics as long as the molecules remain in their lattice sites.

\section{Conclusion}

We have presented a simple method for reshaping a long-range interaction into a short-range one. An interesting extension is to modulate the gradient to generate different types of interaction profiles besides nearest-neighbor ones. For example, one can generate interactions that are still long range but essentially random in sign and magnitude. Such a system would be highly frustrated and would probably be a spin glass \cite{gopalakrishnan11,strack11}.

Another possibility is to engineer spatially anisotropic interactions in a 2D lattice. In our 2D discussion, we assumed that $g_0$ is even, leading to spatially isotropic interactions [Eq.~\eqref{eq:beta_properties}]. If $g_0$ is odd, then the interaction would be anisotropic: $\beta_{r,s}=-\beta_{s,r}$. Anisotropic interactions are known to give rise to exotic physics \cite{bhongale13}.

Lastly, it would be interesting to apply a multifrequency modulation [like Eq.~\eqref{eq:f_multi}] to other Floquet spin models. A multifrequency modulation may be more effective than a single-frequency modulation in terms of realizing topological phases \cite{iadecola15,lee16}.

\section{Acknowledgements}
We thank A.~Gorshkov, Z.-X.~Gong, M.~Foss-Feig, M.~Maghrebi, J.~Young, M.~Lemeshko, and Y.~Yan for useful comments. The numerical optimization was done on Indiana University's supercomputer, Big Red II.

\bibliography{longrange}

\end{document}